\documentclass[aps,pra,twocolumn,halfspacing,10pt,nofootinbib]{revtex4-1}
\usepackage{natbib}
\usepackage[colorlinks]{hyperref}
\usepackage[utf8]{inputenc}
\usepackage[english]{babel}
\usepackage{graphicx}
\usepackage{color}
\usepackage{bm}
\usepackage{amssymb}
\usepackage[intlimits]{amsmath}
\usepackage{amsmath}
\usepackage[para,online,flushleft]{threeparttable}
\newcommand{\pten}[2]{\ensuremath \left( \begin{array}{ccc}
#1 \\[0.3cm]
#2 \\ 
\end{array} \right) }

\begin{document}
\title{Isotope shift, non-linearity of King plots and the search for 
new particles}
\author{V. V. Flambaum$^{1,2}$}
\author{A. J. Geddes$^1$}
\author{A. V. Viatkina$^2$}

\affiliation{$^1$School of Physics, University of New South Wales,
Sydney 2052, Australia}
\affiliation{$^2$Helmholtz Institute Mainz, Johannes Gutenberg University, 55099 Mainz, Germany}

\date{\today}

\begin{abstract}

We derive a mean-field relativistic formula for the isotope shift of an electronic energy level for arbitrary angular momentum; we then use it to predict the spectra of superheavy metastable neutron-rich isotopes belonging to the hypothetical island of stability. Our results may be applied to the search for superheavy atoms in astrophysical 
spectra using the known values of the transition frequencies for the neutron deficient isotopes produced in the laboratory. An example 
of a relevant astrophysical system may be the spectra of the Przybylski's star where superheavy elements up to $Z=99$ have been possibly identified.
%
%
In addition, it has been recently suggested to use the measurements of King plot non-linearity in a search for hypothetical new light bosons. On the other hand, one can find the non-linear corrections to the King-plot arising already in the Standard Model framework. We investigate contributions to the non-linearity arising from relativistic effects in the isotope field-shift, the nuclear polarizability and many-body effects. It is found that the nuclear polarizability contribution can lead to the significant deviation of the King plot 
from linearity. Therefore, the measurements of the non-linearity of King plots may be applied to obtain the nuclear polarizability change between individual isotopes. We then proceed with providing a rough analytical estimate of the non-linearity arising solely from the effect of a hypothetical scalar boson. 
Our predictions give theoretical limitations on the sensitivity of the search for new interactions and should help to identify the most suitable atoms for corresponding experiments.
\end{abstract}

\maketitle

\section{Introduction}

Isotope shift (IS) phenomena in heavy atoms are an important way of probing various scenarios in nuclear physics and can aid the search for new physics beyond the Standard Model. 

Nuclear theory predicts the existence of long-lived isotopes for elements with $Z\geq 104$ (see e.g. \cite{oganessian_heavy_2004,hamilton_search_2013}), in particular isotopes with a magic neutron number $N=184$. However, producing these neutron-rich isotopes in laboratories by colliding lighter atoms is currently impossible. The Coulomb repulsion for nuclei grows as $Z^2$; in order to compensate for this with the attractive strong force, the neutron number $N$ must 
grow faster than $Z$. Consequently, an isotope from the island of stability with $N=184$ cannot be produced from the collision of a pair of lighter isotopes with smaller $N/Z$ ratios.

In contrast to laboratories, various astrophysical events such as supernovae explosions, neutron stars and neutron star - black hole/neutron star mergers generate high neutron fluxes and may create environments favorable for the production of neutron-rich heavy elements. 
For example, a new mechanism of such a kind due to the capture of the neutron star material by a primordial black hole has been suggested in 
\cite{fuller_primordial_2017}. Furthermore, neutron star - neutron star mergers are predicted to generate optimal environments for the production of heavy atoms \cite{goriely_r-process_2011,frebel_formation_2018}. 
As a consequence, astrophysical data may be the best place to observe super-heavy meta-stable elements. It is possible that optical lines of elements up to $Z=99$ have already been identified in the spectra of Przybylski's star \cite{gopka_identification_2008}. These elements include heavy, short-lived isotopes which may be products of the decay of long-lifetime nuclei near the island of stability \cite{dzuba_isotope_2017}. 

IS calculations for superheavy elements can help trace the hypothetical island of stability in existing astrophysical data. It may be possible to predict a spectral line of a neutron-rich isotope $\nu '$ based on the experimental spectrum of a neutron-poor isotope $\nu$ and calculations of IS $\delta \nu$ as $\nu ' = \nu + \delta \nu$. The results can then be used to search for the long-lifetime neutron-rich elements in complicated astrophysical spectra such as that of Przybylski's star.


Spectroscopic measurements of IS may also be relevant to the search for strange-matter. 
Strange nuclei consist of up, down and strange quarks (see \cite{witten_cosmic_1984} and references therein).
A strange-matter nuclei of charge $Z$ would have a very different radius in comparison to any regular isotope. A formula for IS can be used to predict the effects of this change in radius on atomic spectra.

Accurate numerical calculations of IS for heavy and super-heavy elements are usually carried out using sophisticated many-body theory, for example, combining a configuration interaction (CI) and many-body perturbation theory approach (MBPT) (see e.g. \cite{dzuba_isotope_2017} and the references therein). However, in the absence of experimental data a simple analytical formula may be useful for quick estimates of IS and better qualitative understanding of this phenomenon. 
In the present work we derive a relativistic mean-field analytic formula for the field shift, which is the dominating source of IS in heavy atoms. Since the relative magnitude of the many-body corrections to the mean-field case is approximately the same for atoms with similar electronic structure of outer shells, our formula may be used to make reasonable extrapolations from the experimental data of lighter atoms to super-heavy elements where no data are available. 

It should be noted that 
relativistic corrections produce an important difference in the dependence of the field shift on the nuclear radius $r$. The 
traditional expression for field shift is known as $F_i \delta\left<r^2\right>$ where $F_{i}$ is an electronic structure factor and $\delta\left<r^2\right>$ is a nuclear parameter. Instead, the field shift in a relativistic approach should be written as ${\tilde F}_i \delta \left<r^{2 \gamma}\right>$ where $\gamma=\sqrt{(j+1/2)^2 - Z^2 \alpha^2}$, $j$ is the electron angular momentum and $\alpha$ is the fine structure constant; the electronic factor ${\tilde F}_i$ is calculated in the present work.
If one insists on using the traditional formula for the field shift $F_i \delta \left<r^2\right>$, the factor $F_i$  will depend on the nuclear radius, i.e. there will be no factorization of the electron and nuclear variables.

Due to the relativistic effects in heavy atoms, the field shift of the $p_{1/2}$ orbital is comparable to that of the $s_{1/2}$: the ratio is $\sim (1- \gamma)/(1+\gamma)$. The $ Z\alpha$ expansion gives the ratio $\sim Z^2 \alpha^2/4$ but for $Z$=137, $\gamma \approx 0$ and for the superheavy elements the ratio tends to 1. For $j>1/2$ the direct mean-field single-particle field shift is small. However, the mean-field rearrangement effect (the correction to the atomic potential $\delta V$ due to the perturbation of the $s$ and $p_{1/2}$ orbitals by the field-shift operator) 
produces the same dependence of field shift on nuclear radius for all orbitals:
${\tilde F}_i \delta \left<r^{2 \gamma}\right>$, where $\gamma=[1 - Z^2 \alpha^2]^{1/2}$.

Our formula for the field shift allows us to estimate the King-plot nonlinearity of a given element. New long-range forces such as Yukawa-type interactions between electrons and nucleus can lead to nonlinearities in a King plot for a series of isotopes \cite{berengut_probing_2017}. 
It is useful to understand other possible sources of nonlinearities in the IS in order to 
constrain new physics beyond the Standard Model. We estimate the mean-field rearrangement corrections and quadratic effects in the field shift. We also estimate the contribution to IS from the nuclear polarizability which is found to give a bigger contribution to the King-plot non-linearity than the relativistic corrections to the field shift. 
This fact in principle allows an experimental probe of the change of nuclear polarizability between isotopes based on measuring the King-plot non-linearity, 
under the assumption that the effect of possible new physics interactions is negligible.

\section{Field shift in the mean field approximation}
In \cite{racah_isotopic_1932,rosenthal_isotope_1932} the Racah-Rosenthal-Breit formula for IS 
of $s$-wave energy levels was derived using first-order perturbation theory. However it is found that for relativistic cases the formula is unjustified, as it relies on perturbation theory using the Coulomb wave functions for a point-like nucleus when finding the correction to the energy due to the 
finite nuclear size. This is not valid because, while the energy perturbation due to the potential inside the nucleus is small, the perturbed and non-perturbed wave functions within this region are completely different. 
Indeed, the relativistic wave functions for $s_{1/2}$ and $p_{1/2}$ orbitals tend to infinity at $r=0$ while for the finite nucleus they remain finite. 
This problem was already recognized by the authors of the initial publication and since then
numerous attempts have been made to account for this large wave function distortion (see \cite{king_isotope_2013} and references therein). Furthermore, it should be pointed out that a high-precision analytical formula for the finite nuclear size corrections in one-electron atoms and ions was developed, taking into account the fine details of nuclear charge distribution \cite{shabaev_finite_1993}.
However, our goal is to obtain results for many-electron atoms and ions. In this work, we aim to find a simple analytical expression of the field 
shift in many-electron atoms for arbitrary valence electron angular momenta  $j,l$ based on the first-order perturbation theory, but starting from a more realistic initial approximation than a point-like nucleus. 


\subsection{Mean-field isotope shift in many-electron atoms for arbitrary orbital angular momentum}
The dominant contribution to IS in heavy atoms is the field shift arising from the change of nuclear radius, rather then the mass shift which is smaller \cite{sobelman_introduction_1972}.
Let us first consider a model allowing for the estimation of 
field shift for wave functions with arbitrary Dirac quantum numbers $j$ and $l$, where $j=l\pm \frac{1}{2}$. Through this work we assume the nucleus to be a uniformly charged sphere of radius $R$. The nuclear electric potential is:
\begin{equation}\label{eq:v}
 V(r,R)= \begin{cases}
\frac{-Z e^{2}}{r} & \text{for } r \geq R\ , \\
\frac{-Z e^{2}}{R} \left(\frac{3}{2} - \frac{r^{2}}{2 R^{2}} \right) &\text{for } r \leq R\ .
\end{cases}
\end{equation}
In super-heavy nuclei, $V-\frac{Z}{r}$ (the difference between a finite size and point like nucleus) is not a small perturbation (we remind the reader of the collapse of the spectrum for a point like nucleus with $Z=137$). The perturbation used in this work is the change of the potential due to a small relative change of nuclear radii between isotopes which can be defined as 
\begin{equation}\label{eq:dv}
\delta V = \frac{d V(r,R)}{dR}\delta R = \frac{3}{2} \frac{Ze^{2}}{R} \left( 1 - \frac{r^{2}}{R^{2}} \right) \frac{\delta R}{R}.
\end{equation}
Using perturbation theory and integrating over the nucleus
we can find the shift in energy as
\begin{equation}\label{eq:int_kappa}
\delta  E_{\kappa} = \int_{\mathrm{nuc.}} \Psi_{\kappa}^{\dagger} \delta V \Psi_{\kappa} d\vec{r}\ . \\
\end{equation} Radial parts of wave functions can be found from the following Dirac system of radial equations:
\begin{equation}
\begin{cases} 
(\frac{d}{dr}+\frac{\kappa}{r})rf(r)=(m+E-V)rg(r)\ , \\
(\frac{d}{dr}+\frac{\kappa}{r})rg(r)=(m-E+V)rf(r)\ 
\end{cases}
\end{equation}
where $\kappa = \mp \left(j+\frac{1}{2} \right)$, and  $f(r)$ and $g(r)$ are the upper and lower radial components of the Dirac spinor respecitvely. 
We approximate the potential energy 
near $r=0$ to be constant:
\begin{equation}
u=V(0)=-\frac{3Ze^2}{2R},\quad E,m\ll u\ .
\end{equation}
After equating $F=rf(r)$ and $G=rg(r)$ we find that 
\begin{equation}
\begin{cases}
F' +\frac{\kappa}{r}F+uG=0\ , \\
G' -\frac{\kappa}{r}G-uF=0\ .
\end{cases}
\end{equation}
One can check that the solutions at small distances can be written as
\begin{equation*}
\kappa<0\ :\ F = a r ^{|\kappa|} + a_{1} r^{|\kappa|+2}\ ,\quad G=\frac{au}{2|\kappa|+1}r^{|\kappa|+1}\ ,
\end{equation*}
\begin{equation*}
\kappa>0\ :\ G = b r ^{|\kappa|} + b_{1} r^{|\kappa|+2}\ ,\quad F=-\frac{bu}{2|\kappa|+1}r^{|\kappa|+1}\ .
\end{equation*}
where $a$, $b$ are normalization constants and
\begin{equation*}
a_1=-\frac{u^2a}{2(2|\kappa|+1)}\ ,\quad b_1=-\frac{u^2b}{2(2|\kappa|+1)}\ .
\end{equation*}
Here we neglected higher orders in $r^2/R^2$, see also \cite{papoulia_effect_2016}. More accurate calculations have 
demonstrated that their contribution to the field-shift is small (see next subsection).

It can be shown that in both cases 
\begin{equation}
\label{rhoIn}
(F^{2} + G^{2}) \propto r^{2 |\kappa|}\left(1-\frac{9}{2}\frac{Z^{2}\alpha^{2} |\kappa|}{ \left(2|\kappa| +1\right)^{2} } \left(\frac{r}{R}\right)^{2}\right).
\end{equation}
 To determine the field shift we match the expression for the radial density $f^{2} + g^{2}$ inside the nucleus to the radial density outside the nucleus. At the surface $\rho_{inside} = \rho_{outside}$, since $\rho$ is continuous. Near the nucleus the nuclear Coulomb potential is not screened, and all atomic wave functions are proportional to the corresponding Coulomb wave functions. Therefore we use expressions of these wave functions  at small distances (presented in the Appendix C) to approximate the radial density at the nuclear surface ($r=R$):
\begin{align}
\label{rhoOut}
\rho& _{surface} = f_{surface}^{2} +g_{surface}^{2}\nonumber \\
&=\frac{1}{\left(z_{i} + 1 \right)} \frac{Z}{a_B^{3}} \left(\frac{I}{\mathrm{Ry}}\right)^{3/2} \frac{4}{[ \Gamma (2\gamma +1) ]^{2}} \left( \frac{a_B}{2 ZR} \right)^{2-2\gamma} \nonumber   \\ & \times 2 \kappa \left( \kappa - \gamma \right)\ ,
\end{align}
where $\gamma = \sqrt{\kappa^{2} - Z^{2}\alpha^{2}}$, $I = \frac{(z_i+1)^2}{\nu^2}\mathrm{Ry}$ is the ionization energy for an orbital with effective principal quantum number $\nu$  in the ion of charge 
$z_i$,  and $\mathrm{Ry}=\frac{e^{2}}{2 a_B}$ is the Rydberg constant.. As shown above, the electron density inside the nucleus behaves approximately as:
\begin{equation}\label{eq:simpleasym}
\rho(r) \approx \rho_{surface} \; \left(\frac{r}{R}\right)^{2(|\kappa| -1)}.
\end{equation}
This expression approximates the electron density inside the nucleus significantly better than the Coulomb solution and should give more accurate results than the the Racah-Rosenthal-Breit approach. Corrections to Eqs. (\ref{rhoIn}, \ref{rhoOut}, \ref{eq:simpleasym}) are in the next subsection. Their contribution to the isotope shift is small.
 
Using Eq. \eqref{eq:int_kappa} and introducing $x=r/R$ one can find that

\begin{align}
\delta E_{\kappa} =& \frac{3}{2} Z e^{2}R^{2} \frac{\delta R}{R}\int_{0}^{1} \left( f_{\kappa}^{2} + g_{\kappa}^{2} \right) \left(1-x^2 \right ) x^2 dx \nonumber \\
= \frac{3}{2} & Ze^{2}R^{2} \frac{\delta R}{R} \int_{0}^{1} \rho_{sf} x^{2(|\kappa| -1)} \left(1-x^2 \right ) x^2 dx,
\end{align}
which gives
\begin{align}
\delta E_{\kappa}= & \frac{1}{\left(z_{i} + 1 \right)}\frac{12 \kappa (\kappa-\gamma)}{(2|\kappa|+1)(2|\kappa|+3)[ \Gamma (2\gamma +1) ]^{2}}\nonumber \\ &\times \left(\frac{2ZR}{a_B}\right)^{2\gamma}\frac{I^{3/2}}{\mathrm{Ry}^{1/2}}\frac{\delta R}{R}\ .\label{eq:dE}
\end{align}

\subsection{Isotope shift for $s$ and $p_{1/2}$ waves}

From Eq. (\ref{rhoIn}) we see that $Z^2 \alpha^2 r^2/R^2$ corrections to the electron density decrease with the increase of 
$|\kappa|$. Indeed, the ratio of the potential $|V|$ 
and the centrifugal term $|\kappa|/r$ in the Dirac equation decreases as $1/|\kappa|$. Therefore, to analyse the role of the corrections it is sufficient to consider the case of the minimal $|\kappa|=1$, of $s$ ($\kappa=-1$) and $p_{1/2}$ ($\kappa=1$) waves. These are also the most important cases for the isotope shift.

The potential inside the nucleus $V=\frac{-Z e^{2}}{R} \left(\frac{3}{2} - \frac{r^{2}}{2 R^{2}} \right)$ is quadratic and wave functions inside the nucleus correspond to the solutions for the relativistic oscillator. These solutions must be matched with the Coulomb solutions outside the nucleus \footnote{These Coulomb solutions include regular and irregular at $r=0$ components.}. The result may be presented in the following form (see e.g. \cite{khriplovich_parity_1991}):  
\begin{equation}
\Psi_{s_{1/2}}= \pten{f_{s} \Omega_{s}}{i g_{s} \Omega_{p_{1/2}}}\ ,
\end{equation}
\begin{equation}
\Psi_{p_{1/2}}= \pten{-\frac{A_{p}}{A_{s}}\;g_{s} \Omega_{p_{1/2}}}{i \frac{A_{p}}{A_{s}}\;f_{s} \Omega_{s}},
\end{equation}
where $f_{s}$, $g_{s}$, $A_{s}$ and $A_{s}$ are defined in the Appendix A. As before, we treat the change in the potential within the nucleus due to the isotope effect \eqref{eq:dv} as our perturbation.
Again, we find the energy shifts:
\begin{equation}
\delta  E_{s} = \frac{3}{2} Z e^{2}R^{2} \frac{\delta R}{R}\int_{0}^{1} \left( f_{s}^{2} + g_{s}^{2} \right) \left(1-x^2 \right ) x^2 dx\ ,
\label{eq:des}
\end{equation}
\begin{align}
\delta  E_{p_{1/2}} = \frac{3}{2} Z e^{2}R^{2} \left(\frac{A_{p}}{A_{s}}\right)^{2} &\frac{\delta R}{R}\nonumber\\ \times \int_{0}^{1} &\left( g_{s}^{2} + f_{s}^{2} \right) \left(1-x^2 \right ) x^2 dx\ .
\label{eq:dep}
\end{align}
These expressions for the isotope shift are evaluated and expanded over small $Z^{2} \alpha ^{2}$ to give
\begin{align}
\delta  E_{s} =  \frac{1}{z_i+1} \frac{1}{\left[\Gamma(2 \gamma + 1)\right]^{2}} & \left( \frac{2ZR}{a_B} \right) ^{2 \gamma} \frac{I_s^{3/2}}{\mathrm{Ry}^{1/2}}\nonumber\\ &\times \frac{4}{5} (1-0.24 Z^{2}\alpha^{2}) \frac{\delta R}{R}\ ,\label{eq:des2}
\end{align}
\begin{align}
\delta  E_{p} = \frac{1}{z_i+1} \frac{Z^{2}\alpha^{2}}{\left[\Gamma(2 \gamma + 1)\right ]^{2}} &\left( \frac{2ZR}{a_B} \right) ^{2 \gamma} \frac{I_p^{3/2}}{\mathrm{Ry}^{1/2}} \nonumber \\ &\times \frac{1}{5}(1+0.26 Z^{2}\alpha^{2}) \frac{\delta R}{R}\ .\label{eq:dep2}
\end{align}
Up to corrections $\pm 0.01 Z^{2}\alpha^{2}$ these two expressions may be presented as one equation:
\begin{align}\label{eq:kap=1}
\delta  E_{|\kappa|=1} = \frac{4}{5} \frac{1}{z_{i}+1} & \frac{ \kappa \left ( \kappa- \gamma \right )}{\left[\Gamma(2 \gamma + 1)\right]^{2}} \nonumber \\ \times & \left( \frac{2ZR}{a_B} \right) ^{2 \gamma} \frac { I ^{3/2}}{\mathrm{Ry}^{1/2}} \frac {\delta R}{R}.
\end{align}
One can see that the expression \eqref{eq:kap=1} directly follows from \eqref{eq:dE}, if we put $|\kappa|=1$. Estimates show that the higher order correction $\sim  Z^{4}\alpha^{4}$  comes with a small coefficient. 

Note that the ratio of the isotope shifts for $p_{1/2}$ and $s_{1/2}$ is equal to 
\begin{equation*}
\left( \frac{A_{p}}{A_{s}}\right)^{2} =\left( \frac{I_{p}}{I_{s}}\right)^{3/2} \frac{z^{2} \alpha^{2}}{4} \left(1+\frac{z^{2} \alpha^{2}}{4}  \right)^2\approx \left(\frac{I_{p}}{I_{s}}\right)^{3/2}\frac{1-\gamma}{1+\gamma}
\end{equation*}

\begin{table*}[t]
\caption{\label{table:shift} Estimates of the isotope shift  $\delta \nu$ (using  Eq.\eqref{eq:dE}) for 
a given transition in superheavy atoms. $A_1$ is the atomic number of already synthesised reference isotope. $A_2=Z+184$ is the isotope of a given element belonging to the hypothetical island of stability with magic neutron number $N=184$.}~\\[-0.5cm]
\begin{tabular}
{p{1.8cm} p{1.8cm} p{1.8cm} p{1.8cm} p{1.5cm} p{0.7cm} p{2.2cm} p{2cm} p{2cm}}
 \multicolumn{9}{c}{} \\
 \hline \hline & &  \\ 
 &Atom & &  & Transition & & &
 $\delta \nu$ (cm$^{-1}$) & $\delta\nu$ (GHz)\\ [0.3cm]
Symbol & Z & A$_{1}$ & A$_{2}$ &   &  \\
\hline
& &  \\ [-0.5cm]
Cf     & 98  & 251 & 282 & $ 5f^{10} 7s^{2}$ & $-$ & $5f^{10} 7s7p$ & -7.3 & -218\\
Es     & 99  & 252 & 283 & $ 5f^{11} 7s^{2}$ & $-$ & $5f^{11} 7s7p$ & -7.8 & -233\\
Fm     & 100 & 257 & 284 & $ 5f^{12} 7s^{2}$ & $-$ & $5f^{12} 7s7p$& -7.7 & -230\\
Md     & 101 & 258 & 285 & $ 5f^{13} 7s^{2}$ & $-$ & $5f^{13} 7s7p$& -8.5 & -255\\
No     & 102 & 259 & 286 & $ 7s^{2} $ & $-$ & $7s7p$ & -9.6 & -286\\
Lr     & 103 & 266 & 287 & $ 7s^{2}7p$ & $-$ & $7s^{2}8s $ & 0.78 & 23.3\\
Rf     & 104 & 263 & 288 & $ 6d^{2} 7s^{2}$ & $-$ & $6d^{2} 7s7p $ &-11.5 & -344\\
Db     & 105 & 268 & 289 & $ 6d^{3} 7s^{2}$ & $-$ & $6d^{3} 7s7p $ &-11.7 & -351\\
Sg     & 106 & 269 & 290 & $ 6d^{4} 7s^{2}$ & $-$ & $6d^{4} 7s7p $ &-14.1 & -424\\
Bh     & 107 & 270 & 291 & $ 6d^{5} 7s^{2}$ & $-$ & $6d^{5} 7s7p $ &-17.1 & -511\\
Hs     & 108 & 269 & 292 & $ 6d^{6} 7s^{2}$ & $-$ & $6d^{6} 7s7p $ &-22.3 & -670\\
Mt     & 109 & 278 & 293 & $ 6d^{7} 7s^{2}$ & $-$ & $6d^{7} 7s7p $ &-17.3 & -518\\
Ds     & 110 & 281 & 294 & $ 6d^{8} 7s^{2}$ & $-$ & $6d^{8} 7s7p $ &-17.8 & -533\\
Rg     & 111 & 282 & 295 & $ 6d^{9} 7s^{2}$ & $-$ & $6d^{9} 7s7p $ &-21.1 & -632\\
Cn     & 112 & 285 & 296 & $ 6d^{10} 7s^{2}$ & $-$ & $6d^{10} 7s7p $ &-21.1 & -633\\
Nh     & 113 & 286 & 297 & $7s^{2}7p $ & $-$ & $7s^{2}8s$ & -0.35 & -10.5\\
Fl     & 114 & 292 & 298 & $7p^{2}$ & $-$ & $7p8s$ & -0.64 & -19.3\\
\hline \hline 
\end{tabular}

\end{table*}

\section{
Quantitative 
Field Shift Estimates 
}
\subsection{Estimates for Field Shifts in Superheavy Atoms}
Table \ref{table:shift} depicts the estimates for isotope shift in superheavy atoms which were calculated using 
Eq. \eqref{eq:kap=1}. The ionization potentials used to calculate the field shift for each level in a given atom has been detailed in Appendix D. Furthermore, the nuclear radius $R$ was found using $R= r_{0} A^{1/3}$ where we assumed that 
$r_{0} = 1.15$ fm for the purpose of these calculations. 

One of the motivations for the current work was to provide a simple method to estimate IS which is suitable for superheavy atoms and provides a better understanding of its dependence on the nuclear and atomic parameters. Accurate many-body calculations of the field shift for No, Lr, Nh, Fl and Ra have recently been 
performed and presented in \cite{dzuba_isotope_2017}. The CI+MBPT isotopic shift value for No was found to be -7.28 cm$^{-1}$. Our approximate IS value for No is -9.6 cm$^{-1}$ as presented in \ref{table:shift}. The difference  is actually comparable  to  20\% error of the CI+MBPT value.

The calculated IS for Lr, Nh and Fl is small due to large cancellations in the shifts between the lower $p$ state and excited $s$ state. We hence can provide only an order of magnitude estimates when calculating IS for transitions for $ p \rightarrow s$ states using the method presented in this paper.  Indeed, the $7p_{1/2}$ state IS is suppressed by the factor $(1-\gamma)/(1+  \gamma)$ but enhanced by the higher $7p_{1/2}$ ionization potential than that of $8s$. This is why IS of $7p_{1/2}$ and $8s$ states nearly cancel each other out. While the absolute accuracy of IS calculations is the same, the relative accuracy of IS of  the transition energy  is poor.
The value for the overall IS of Nh in our case is -0.35 cm$^{-1}$ which is significantly smaller than and opposite in sign to the CI+MBPT value for Nh which was stated to be 1.42 cm$^{-1}$. Similarly, we calculated an IS of -0.64 cm$^{-1}$ for Fl which is the same order of magnitude as CI+MBPT value given as 0.12 cm$^{-1}$ yet also opposite in sign.
 Our approximate value of IS for Lr is 0.78 cm$^{-1}$ which is notably smaller then the CI+MBPT value of 3.134 cm$^{-1}$. 
 We re-iterate that the relative accuracies of the analytical formula and the CI+MBPT  method in these cases are low, and all what we can conclude is that the IS is small and the frequencies of the transitions in all isotopes will be approximately the same. 

\subsection{Estimates for Field Shifts in Ca, Ca$^+$, Yb and Hg}
We calculated field shifts of $s\rightarrow p$ transitions for 
Ca, Ca${^+}$, Yb and Hg 
to compare with known experimental data. The results are presented in Tables \ref{table:experiment1} and \ref{table:experiment2}. The agreement for $s-p$ transitions is good. However, in the case of Ca$^+$, formula \eqref{eq:dE} underestimates the measured field shifts of $3p^{6}3d\;^{2}D_{3/2} \rightarrow 3p^{6} 4p \;^{2}P_{1/2}$ transition \cite{gebert_precision_2015} by two orders of magnitude. The reason is that the direct field IS in $d-p$ transitions in light atoms  is very small and the actual field IS  is dominated by the mean-field rearrangement effect (the change of atomic potential due to the isotope shift in $s$ and $p_{1/2}$ wave functions) which will be discussed in the next section.

\begin{table}[hb]
\caption{Comparison of experimental field shifts\label{table:experiment1} in Ca, Yb and Hg with theoretical prediction based on formula \eqref{eq:dE} and experimental values of mean nuclear charge radii. Both measured field shifts and nuclear charge radii are found in \cite{fricke_nuclear_2004}.}
\begin{tabular}
{p{0.85cm} p{0.5cm} p{0.5cm} p{2.8cm} p{2.0cm} p{1.3cm}}
 \hline \hline & &  \\
Atom &$A_1$&$A_2$ &Transition & $\delta\nu_\mathrm{exper}$ (MHz) & $\delta\nu_\mathrm{theor}$ (MHz)\\
\hline
Ca & 46 & 48 & $3p^6 4s^2 - 3p^6 4s 4p$ & $-25.3\pm 1.0$ & $-31$\\
Yb & 174& 176& $4f^{14}6s^2 - 4f^{14}6s6p$ & $\:993\pm 250$ & $1217$ \\ 
Hg & 202& 204& $5d^{10}6s^2 - 6d^{10}6s6p$ & $5238\pm 11$ & $4939$ \\
\hline\hline
\end{tabular}
\end{table}

\begin{table}[htb]
\caption{\label{table:experiment2} Comparison of experimental field shifts for the $3p^{6}4s\;\;^{2}S_{1/2}\rightarrow 3p^{6} 4p \;\;^{2}P_{1/2}$   transition in Ca$^{+}$ \cite{gebert_precision_2015} with theoretical results using formula \eqref{eq:dE} and nuclear charge radii data from \cite{fricke_nuclear_2004,ruiz_unexpectedly_2016}.
}~\\[-0.5cm]
\begin{threeparttable}[htb]
\begin{tabular}
{ p{1.5cm} p{2.2cm} p{2.2cm} p{2cm} }
\multicolumn{4}{c}{} \\
\hline \hline & &  \\ 
& & MHz & \\
& $\delta \nu_{field} ^{40, \;42}$& $\delta \nu_{field} ^{40, \;44}$& $\delta \nu_{field} ^{40, \;48}$\\
\hline

Theory            & -60.1 & -85.2 & -0.30\\
Exp.         & -60.9(2.0) & -79.6(2.7) & 1.27(1.69)\\ 
\hline \hline 
\end{tabular}

 \end{threeparttable}

 \end{table}

\section{Non-linearities in King Plot for isotope shifts}
As we will show below the non-linear corrections to the King plot may be due to the non-factorization of the electronic and nuclear parameters in the expression for the field IS. This non-factorization appears if we have two or more nuclear parameters which are not proportional to each other and appear in different combinations for different atomic transitions. As a rule, the second nuclear parameter  gives a contribution to IS which is several orders of magnitude smaller than the contribution of the first nuclear parameter. This means that the non-linearity is small. 

\subsection{King plot in the mean-field approximation} 
With $\mu_{AA'} = \frac{1}{m_{A}} - \frac{1}{m_{A'}}$, where the masses of isotopes $A$ and $A'$ are denoted as $m_{A}$ and $m_{A'}$ respectively, the IS can be written as follows:
\begin{equation}
\nu^{AA'}_{i} = K_{i}\mu_{AA'} + F_{i} \delta \left < r^{2 \gamma_{1}} \right >_{AA'} + G_{i} \delta \left< r^{2 \gamma_{2}} \right>_{AA'}\ . \label{eq:mu}
\end{equation}
The first term expresses the mass shift, both normal and specific. 
The second term is the leading-order contribution to the field shift discussed in the previous section. For $s$-wave in the non-relativistic limit ($\gamma_1\rightarrow 1$)
this term scales with the difference of mean squares of nuclear charge radii between the isotopes $A$ and $A'$: $\delta \left < r^{2} \right >_{AA'}$. The third term expresses a correction to the field shift produced by  higher waves. For  example, for $p_{3/2}$ in  the non-relativistic limit, $\gamma_2\rightarrow 2$ and $\delta \left < r^{2 \gamma_{2}} \right >_{AA'} = \delta \left < r^{4} \right >_{AA'}$.
A similar term may appear as a sub-leading correction to the $s$-wave field shift  but it does not produce a contribution to the non-linearity of the King plot since it only redefines the the main $s$-wave contribution $F_{i} \delta \left < r^{2 \gamma_{1}} \right >_{AA'}$, i.e. it produces a correction which is the same in different atomic transitions. The non-linearity appears when the ratio $G_i/F_i$ changes, 
as will be shown below.

\begin{figure}[b]
\includegraphics[width=\columnwidth]{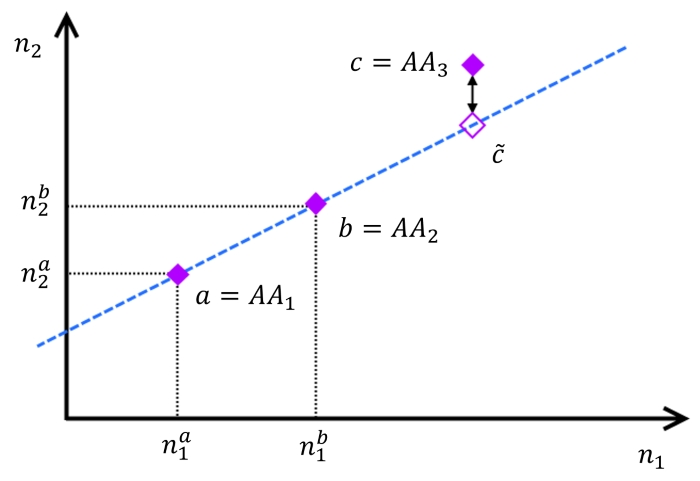}
\caption{Schematic illustration of the King plot non-linearity. Modified frequencies $n_1$ and $n_2$ are plotted for the three pairs of isotopes \eqref{eq:isotopes}. The difference between point $c$ and the hypothetical point $\tilde{c}$ lying on the same line as $a$ and $b$ leads to the evaluation of non-linearity of the plot \eqref{eq:nonlin_def}.}\label{fig:King}
\end{figure}

Next we divide equation \eqref{eq:mu}  by $ \mu_{AA'}$ and hence define a new modified frequency $n_{i}=\nu^{AA'}_{i}/\mu_{AA'}$ :
\begin{equation}
n_{i}=K_{i} + F_{i}x + G_{i}y\ ,
\end{equation}
with $x = \delta \left < r^{2 \gamma_{1}} \right >_{AA'} / \mu_{AA'}$ and $y=\delta \left< r^{2 \gamma_{2}} \right>_{AA'} / \mu_{AA'}$. Here $i=1, 2$ are the two transitions examined in a chain of isotopes.
A plot of $n_{1}$ vs. $n_{2}$ gives what is known as 
King plot \cite{king_comments_1963}.  If we assume $G_{1,2}=0$, one can write $n_2$ as a linear function of $n_1$, i.e. the King plot is linear. To trace the possible non-linearity, we must consider at least 4 isotopes ($A, A_1, A_2, A_3$) forming 3 pairs, giving three points on the plot:
\begin{equation}
AA_1\equiv a,\quad AA_2\equiv b,\quad AA_3\equiv c\ .\label{eq:isotopes}
\end{equation}
The first two points can be used to determine the gradient $k=(n_2^b-n_2^a)/(n_1^b-n_1^a)$. Let us state a hypothetical $\tilde{c}$ point, lying on the same line as $a$ and $b$:
\begin{equation}
n_2^{\tilde{c}}= n^b_2 + k(n_1^c-n_1^b)\ .
\end{equation}
Then the non-linearity is defined (see Fig. \ref{fig:King}):
\begin{equation}\label{eq:nonlin_def}
\mathrm{NL}\equiv (n^c_2-n_2^{\tilde{c}})\mu_{c}\ ,
\end{equation}
where $\mu_c=\mu_{AA_3}$, and the value \eqref{eq:nonlin_def} can be expressed in Hz. Let us explicitly expand the difference between two modified frequencies:
\begin{align*}
&n^c_2-n_2^{\tilde{c}} = K_2+F_2 x^c+G_2 y^c - K_2-F_2x^b-G_2y^b\\
-&\frac{F_2(x_b-x_a)+G_2(y_b-y_a)}{F_1(x_b-x_a)+G_1(y_b-y_a)}[F_1(x_c-x_b)+G_1(y_c-y_b)] .
\end{align*}
Under the assumption of $\frac{G\Delta y}{F\Delta x}\ll 1$ and with $q_{ba}\equiv\frac{y_b-y_a}{x_b-x_a}$, one can show that:
\begin{equation}
n^c_2-\tilde{n}_2^c=(x_c-x_b)(q_{cb}-q_{ba})\left(\frac{G_2}{F_2}-\frac{G_1}{F_1}\right)F_2 .
\end{equation}
Here we see that the non-linear correction vanishes in two cases: \begin{enumerate}
\item $q_{cb}=q_{ba}$,
\item $\frac{G_2}{F_2}=\frac{G_1}{F_1}$. 
\end{enumerate}
The first case considers $\delta \left<r^{2\gamma_1}\right>$ and $\delta\left<r^{2\gamma_2}\right>$. These parameters are correlated: generally speaking, increase of the nuclear radius $R$ leads to the increase of both $\delta \left<r^{2\gamma_1}\right>$ and $\delta\left<r^{2\gamma_2}\right>$. 
It is easy to check that if the field shift for all isotopes is completely defined by the change of the nuclear radius $\delta R$ from isotope to isotope, i.e. if $\delta \left<r^{2\gamma}\right> =D_{\gamma}\delta R$ for any $\delta R$ (as in the linear approximation in $\delta R$), we have $q_{cb}=q_{ba}$. 

For example, in the isotope shift \eqref{eq:dE} the dependence on $R$ is given by $R^{2\gamma-1}\delta R$ and the difference between two isotopes is proportional to $R_A^{2\gamma-1}\delta R_{A\tilde{A}}$. This, given fixed reference isotope A, leads to $q_{cb}=q_{ba}$. Therefore, to get a non-zero result we should go beyond the first order in $\delta R/R$. We integrate formula \eqref{eq:dE} to effectively include all orders of perturbation theory in $\delta R/R$. Then the field shift of an energy level between isotopes $A$ and $\tilde{A}$ is:
\begin{align}
\Delta E_{\kappa}= &\frac{12 \kappa (\kappa-\gamma)}{2\gamma (z_i+1) (2|\kappa|+1)(2|\kappa|+3)[ \Gamma (2\gamma +1) ]^{2}}\nonumber \\ \times 
&\left(\frac{2Z}{a_B}\right)^{2\gamma}\frac{I^{3/2}}{\mathrm{Ry}^{1/2}} \left(R_{\tilde{A}}^{2\gamma}-R_A^{2\gamma}\right)\ .\label{eq:DE}
\end{align}
We plot the field shifts of transition frequencies $(\Delta E_{\kappa,\mathrm{upper}}-\Delta E_{\kappa,\mathrm{lower}})_{A\tilde{A}}$ for three pairs of isotopes of four elements: Ca$^+$, Sr$^+$, Yb$^+$ and Hg$^+$. Fitting a line to the first two pairs $a$ and $b$, we find the nonlinearity as $(n^c_2-\tilde{n}_2^c)\mu_c$. 

The results are presented in Table \ref{table:king}.
In order to find the radii $R$ for substituting into \eqref{eq:DE}, we first make use of the simple liquid-drop model where $R=r_0 A^\frac{1}{3}$ (Method 1 in Table \ref{table:king}). Then to obtain more realistic estimate we find the equivalent $R$ from the experimental values of mean square nuclear charge radius \cite{fricke_nuclear_2004,ruiz_unexpectedly_2016} (Method 2 in Table \ref{table:king}): $R^2=\frac{5}{3}\left<r^2 \right>$. 

The expression \eqref{eq:DE} is, in fact, only the first-order contribution to the field shift in terms of energy. The second order in the single-electron mean-field approximation can be roughly estimated as a quadratic term $\pm (\Delta E_{\kappa})^2/I$. Due to its smallness in higher waves ($\kappa\neq -1$) it has only a negligible effect on the King plot non-linearity. But, as will be shown in Sec. \ref{sec:MB}, if we include many-body  corrections, this quadratic term can alter considerably the non-linearity value in heavy atoms (the field shift is $ \propto Z^{2 \gamma}$, correspondingly the quadratic term 
should be $ \propto Z^{4 \gamma}$).

\subsection{Nuclear polarizability effect}
The nuclear structure effects for simple atoms have been considered in Ref. \cite{puchalski_nuclear_2010}. We are interested in such effects in many-electron atoms.  
The nuclear polarization potential produced by the  nuclear polarizability $\alpha_p$ is a long-range one ($V_{\alpha}=-\frac{1}{2}\frac{\alpha_p e^2}{r^4}$, $\alpha_p$ having dimension $[l^3]$), therefore it can give a significant contribution to a higher-wave isotope shift and overall non-linearity of King plot. The main contribution to the corresponding energy shift comes from the area near  the nucleus where the nuclear potential is not screened. In order to estimate the shift, we integrate the radial density $\rho_\kappa=f^2_\kappa + g^2_\kappa$ using atomic wave functions proportional to the  Coulomb wave functions outside the nucleus (see Appendix B) with the  interaction Hamiltonian $V_{\alpha}$. 
\begin{equation}\label{eq:dEe}
\delta E_{\alpha} = \int_{r_0}^{+\infty}\left(f^2_\kappa(r)+g^2_\kappa(r)\right)\left(-\frac{1}{2}\frac{\alpha_p e^2}{r^4}\right)r^2dr\ ,
\end{equation}
\begin{equation*}
r_0 = \begin{cases}
R &,\ |\kappa|=1\ , \\
0 &,\ |\kappa|>1\ .
\end{cases}
\end{equation*}
 For $|\kappa|>1$ the integral with the Coulomb wave functions converges at $r=0$ and we may calculate it taking the cut-off parameter $r_0=0$. In this case the general analytic solution 
 can be presented here as:
\begin{align}
\delta E_{\alpha} = -\alpha_p\frac{9+5\kappa(\kappa-3)+5Z^2\alpha^2+\gamma^2}{256\gamma(-9+\gamma^2(7-4\gamma^2)^2)}&\nonumber\\ \times\frac{8^3 Z^2}{a_B^3}\frac{2}{z_i+1}&\frac{I^{3/2}}{\mathrm{Ry}^{1/2}}\ , \label{eq:dEa_kappa}
\end{align}
where $\gamma=\sqrt{\kappa^2-Z^2\alpha^2}$ and $I$ is the ionization potential of the electron, $a_B$ the Bohr radius, $\mathrm{Ry}$ the Rydberg constant and $z_i$ the ion charge.
For $|\kappa|=1$, the  integral in Eq.~\eqref{eq:dEe} from 0 would diverge, but a cut-off from the nuclear radius $R$ would give a reasonable upper estimate of the effect.\footnote{Actually, the polarization potential $V_{\alpha}$ becomes a non-local integration operator \cite{dzuba_correlation_1987} starting from larger distances, $r < r_0 \sim 10$ fm,  where the relativistic kinetic energy of electron $\sim \hbar c/r_0$ 
approaches $E \approx 20$~MeV. Here $E$ is the excitation energy of the nuclear giant dipole resonance which gives the dominating contribution to the polarizability. This problem will be discussed in a future publication.}  
Note that $s$ orbital always appears in both transitions which we compare in the King plot, therefore the  exact magnitude of this $|\kappa|=1$ term is not important for the estimate of the non-linearity.

To obtain the numerical values for the integral with $|\kappa|=1$ we have used the Bessel function solutions from Appendix B, however, without loss of the actual numerical accuracy its adequate approximation can be found by substituting expressions for  wave functions expanded at  $r\rightarrow 0$ (see Appendix C) to the integral \eqref{eq:dEe}:
\begin{align}
\delta E_{\alpha}=-\alpha_p &\frac{8\kappa(\kappa-\gamma)}{3-2\gamma}\frac{1}{z_i+1}\frac{Z}{a^2_B}\ \times \nonumber \\
&\frac{1}{[\Gamma(2\gamma+1)]^2}\left(\frac{a_B}{2 Z}\right)^{2-2\gamma}\frac{I^{3/2}}{\mathrm{Ry}^{1/2}}R^{2\gamma-3}\ . \label{eq:dEa_1}
\end{align}
Although the ceiling estimate presented in equation \eqref{eq:dEa_kappa} is utilized in subsequent calculations and is sufficient for the purpose of this work, an expression can be written specifically for the s$_{1/2}$ contributions to nuclear polarizability. The non-relativistic expression for the energy shift has been derived in \cite{pachucki_theory_1994}: 

\begin{equation}
\delta E_{\alpha, s} =  - m_e c^2 \alpha \Psi (0)^{2} \alpha_p \left [ \frac{19}{6} + 5 \ln \left ( \frac{\bar{E}}{m_e c^2} \right ) \right] \label{eq:nonrel}
\end{equation}
where $\bar{E}$ is the average nuclear excitation energy. Using the the nuclear oscillator model we can estimate the excitation energy as the distance between the nuclear shells $\bar{E} = 40\ \textrm{MeV}/A^{1/3}$. $\Psi (0) ^ 2$ is the single valence electron density at the nucleus which, in the non-relativistic limit has the form 
of 

\begin{equation}
\Psi (0) ^ 2 = \frac{Z \; \left (\frac{I}{Ry}\right )^{3/2}}{\pi a_B^3 (z_i + 1)}.
\label{eq:valElecDens}
\end{equation}

The expression for $\delta E_{\alpha, s}$ can be modified to include a relativistic factor that reflects the increase in the relativistic wavefunction towards the nucleus. To make a rough estimate of this factor, we chose the cut off radius to be the Compton wavelength of an electron $k_e = \frac{\hbar}{m_e c}$ where the non-relativistic approach breaks down:

\begin{equation}
R_{rel} = \left(\frac{2 Z k_e}{a_B}\right)^{2\gamma -2} \approx \left( \frac{1}{Z \alpha} \right)^{2(1-\gamma)}
\end{equation}
Incorporating $R_{rel}$ with the non-relativistic expression for $\delta E_{\alpha, s}$ we can present the expression

\begin{align}
\delta E_{\alpha, s} =  - m_e c^2 \alpha & \frac{Z \; }{\pi a_b^3 (z_i+1) }  \left (\frac{I}{Ry}\right )^{3/2}\alpha_p  \\ & \left [ \frac{19}{6} + 5 \ln \left ( \frac{\bar{E}}{m_e c^2} \right ) \right]R_{rel}.
\end{align}

The relativistic factor can be written as $R_{rel} \approx (Z \alpha)^{-Z^2 \alpha^2} $ by expanding $\gamma$ through by small $Z \alpha$. It follows that $R_{rel} \approx e^{\ln((Z \alpha)^{-Z^2 \alpha^2})} $ can be approximated as $R_{rel} \approx 1 + Z^2 \alpha^2 \ln(\frac{1}{Z \alpha})$. Hence it is apparent that the relativistic formulation of the $s_{1/2}$ contribution to nuclear polarizability differs from Eqs. \eqref{eq:nonrel} - \eqref{eq:valElecDens} by less then a factor of 2.

An expression for nuclear polarizability $\alpha_{E}$ based on the giant resonance approach  was obtained by Migdal \cite{migdal_quadrupole_1945,levinger_migdals_1957}: 
\begin{equation}\label{eq:alpha_migdal}
\alpha_{p}=\frac{e^2 R^2 A}{40\ a_{sym}}\ . 
\end{equation}
We use the empirical value of the nuclear symmetry energy $a_{sym}=23$~MeV \cite{von_weizsacker_theory_1935,green_coulomb_1954}. 

To the first order one can treat the change of $\alpha_E$ as consisting of two independent parts, one resulting from the growth of nuclear radius $\Delta R$ and another from the change of nucleon number $\Delta A$:
\begin{equation}
\Delta\alpha_p=\frac{2e^2 R}{40\ a_{sym}} A\Delta R+\frac{e^2 R^2}{40\ a_{sym}} \Delta A\ .\label{eq:Da}
\end{equation}
The second contribution can be independently evaluated using the nuclear oscillator model. From the second-order perturbation theory follows that with the addition of one neutron, the nuclear polarizability changes as:
\begin{equation}
\delta\alpha_{p,\mathrm{A}+1}=-q_n^2\left[\frac{\langle n|x|n+1\rangle^2}{E_n-E_{n+1}}+\frac{\langle n|x|n-1\rangle^2}{E_n-E_{n-1}} \right]
\end{equation}
Here $q_n=eZ/A$ is the effective charge of a neutron, originating from the recoil effect
\footnote{The second term with the matrix element $\langle n|x|n-1\rangle$ emerges from the single-particle consideration. In the many-body language it is the "blocking" contribution: core neutrons cannot be excited to the state occupied by the valence neutron.}. Energy levels of a quantum oscillator are known to be $E_n=\hbar\omega(n+\frac{1}{2})$ and its matrix elements can be written \cite{landau_quantum_1958}:
\begin{equation*}
\langle n|x|n+1\rangle^2=\frac{(n+1)\hbar}{2 M\omega}\ ,\quad \langle n|x|n-1\rangle^2=\frac{n\hbar}{2 M\omega}\ ,
\end{equation*}
with the frequency $\omega$ for the case of nuclei and $M$ being the neutron mass. Assuming $r_0=1.15$~fm, one can write:
\begin{equation*}
\omega=\frac{40\ \mathrm{MeV}}{\hbar}\frac{r_0}{R}\ ,
\end{equation*}
\begin{align}
\Delta\alpha_{p,A}=&\delta\alpha_{p,\mathrm{A}+1}\Delta A=\frac{e^2}{2M\omega^2}\left(\frac{Z}{A}\right)^2\Delta A\nonumber\\
=&\frac{e^2\hbar^2}{2M}\left(\frac{Z}{A}\right)^2\left(\frac{R}{r_0\times(40\ \mathrm{MeV})}\right)^2\Delta A .\label{eq:DaA}
\end{align}
The second term in \eqref{eq:Da} and expression \eqref{eq:DaA} are close in value and they both depend on $R^2$. It means that the formula \eqref{eq:alpha_migdal} effectively includes the contribution from adding neutrons and we can use it alone to estimate the change of nuclear polarizability between isotopes. We introduce the empirical coefficient $\zeta(A)$ to scale our prediction to the more accurately evaluated nuclear polarizabilities in \cite{piekarewicz_electric_2012}.
\begin{equation*}
\zeta(A)=0.76+\frac{2.79}{A^{1/3}}\ ,
\end{equation*}
\begin{equation}
\alpha_p=\zeta(A)\frac{e^2 R^2 A}{40\ a_{sym}}\ .
\end{equation}
This final expression of nuclear polarizability is used to model the King plot non-linearity, which grows dramatically compared to the non-linearity found only according to the field shift formula \eqref{eq:DE}, as can be seen in Table \ref{table:king}.

\begin{table*}[t]
\caption{\label{table:king}Estimates for the non-linearities of King plot (defined in \eqref{eq:nonlin_def}). Methods 1 and 2 are based on the mean-field analytic expression of field isotope shift \eqref{eq:DE} found from the estimate of wave function density \eqref{eq:simpleasym} in an uniformly charged spherical nucleus. Method~1 utilizes the liquid drop approximation for nuclear radius $R=r_0 A^\frac{1}{3}$ with $r_0=1.15$~fm and Method~2 uses experimental data \cite{fricke_nuclear_2004,ruiz_unexpectedly_2016} for mean squares of nuclear charge radii $\left< r^{2} \right>$ when finding the equivalent nuclear radius: $\left< r^{2} \right>=\frac{3}{5}R^{2}\ $. Method~3 accounts for both expression \eqref{eq:DE} and the contribution of nuclear polarizability \eqref{eq:dEe}, equivalent nuclear radius $R$ is again based on the experimental data, i.e. it is the most complete calculation in this table. 
We have not calculated these corrections and the nuclear polarizability contribution in $s-p/d-p$ transitions in Ca$^+$ since they are expected to be similar to  $s-d$ transitions.
}~\\[-0.5cm]
 \begin{threeparttable}
\begin{tabular}
{p{0.8cm} p{0.5cm} p{0.5cm} p{0.5cm} p{0.5cm} p{0.8cm} p{4.9cm} p{2.7cm} p{2.5cm} p{2.5cm}
}
 \multicolumn{7}{c}{} \\
 \hline \hline & &  \\ 
 Ion & & & & & & Pair of transitions & Non-linearity (Hz)\\ [0.3cm]
  & Z & A & A$_{1} $& A$_{2}$ & A$_{3}$ & & Method 1 & Method 2 & Method 3\\
\hline
& &  \\ [-0.5cm]
Ca$^{+}$     & 20  & 40 & 42 & 44 & 48 & $3p^{6}4s\;\;^{2}S_{1/2}\rightarrow 3p^{6} 4p \;\;^{2}P_{1/2}$ &$-1.2 \times 10^{-4}$ & $3.0 \times 10^{-4}$ & -- \\
& & & & & & $3p^{6}3d\;\;^{2}D_{3/2} \rightarrow 3p^{6} 4p \;\;^{2}P_{1/2}$ & & \\
\hline 
& & & & & & $3p^{6}4s\;\;^{2}S_{1/2}\rightarrow 3p^{6}3d\;\;^{2}D_{3/2}$ &$-1.2 \times 10^{-4}$ & $3.0 \times 10^{-4}$ & $-6.6\times 10^{-2}$ \\
& & & & & & $3p^{6}4s\;\;^{2}S_{1/2}\rightarrow 3p^{6}3d\;\;^{2}D_{5/2}$ & & &\\
\hline
Sr$^{+}$  & 38  & 84 & 86 & 88 & 90 &  $4p^{6}5s\;\;^{2}S_{1/2}\rightarrow 4p^{6} 4d\;\; ^{2}D_{3/2}$ & $-1.1\times 10^{-3}$ & $1.1 \times 10^{-2}$ & $-2.6$ \\
& & & & & & $4p^{6}5s\;\;^{2}S_{1/2}\rightarrow 4p^{6} 4d\;\;^{2}D_{5/2}$ & & \\
\hline
Ba$^{+}$ & 56 & 132 & 134 & 136 & 138 & $ 5 p^6 6 s^1 \;\;^{2}S_{1/2}\rightarrow 5 p^6 5 d \;\;^{2}D_{3/2}$ & $-3.6 \times 10^{-3}$ & $ -3.9 \times 10^{-2}$ & 7.6 \\
& & & & & & $ 5 p^6 6 s^1 \;\;^{2}S_{1/2}\rightarrow 5 p^6 5 d \;\;^{2}D_{5/2}$ & & 
\\
\hline
Yb$^{+}$ & 70 & 168 & 170 & 172 & 176 & $4f^{14}6s\;\;^{2}S_{1/2}\rightarrow 4f^{13}6s^{2}\;\;^{2}F_{7/2}^{o}$ & $6.1\times 10^{-2}$ & $-3.1$ & 38 \\ 
& & & & & & $4f^{14}6s\;\;^{2}S_{1/2}\rightarrow 4f^{14}5d\;\;^{2}D_{3/2}$ & & \\
\hline
& & & & & & $4f^{14}6s\;\;^{2}S_{1/2}\rightarrow 4f^{14}5d\;\;^{2}D_{3/2}$ & $-6.1\times 10^{-2}$ & $3.1$ & $-18$ \\
& & & & & & $4f^{14}6s\;\;^{2}S_{1/2}\rightarrow 4f^{14}5d\;\;^{2}D_{5/2}$ & & \\
\hline
Hg$^{+}$     & 80  & 196 & 198 & 200 & 204 &  $5d^{10}6s\;\; ^{2}S_{1/2}\rightarrow 5d^{9}6s^{2}\;\;^{2}D_{3/2}$ & $5.5\times 10^{-1}$ & $3.1$ & $-14$\\
& & & & & & $5d^{10}6s\;\;^{2}S_{1/2}\rightarrow 5d^{9}6s^{2}\;\;^{2}D_{5/2}$\\
\hline \hline 
\end{tabular}

\end{threeparttable}

\end{table*}

\begin{table*}[t]

\caption{\label{table:kingMB} Estimates for the non-linearities of King plot (defined in \eqref{eq:nonlin_def}), taking into account the many-body mean-field rearrangement corrections, as shown in \eqref{eq:kappa_mb1}. Method 4 is based on a sum of the mean-field analytically determined contribution \eqref{eq:DE} and the first-order many-body effect $(-\Delta \varepsilon_\kappa/2)$. Method 5 uses the mean-field term \eqref{eq:DE}, the first-order many-body effect and the polarizability contribution \eqref{eq:dEe}. The next two columns show the second-order contributions to the non-linearity with an unknown sign arising from the quadratic terms (e.g. \eqref{eq:s_mb2} and \eqref{eq:kappa_mb2}). The first of them shows the quadratic corrections ignoring the nuclear polarizability $\alpha_p$, i.e. it corresponds to Method 4. The second of them takes into account the polarizability contribution and thus corresponds to Method 5. The very last column details the nonlinearity due to the quadratic mass shift (QMS) estimated from the normal mass shift contribution as presented in \eqref{eq:M2}.
}~\\[-0.5cm]
 \begin{threeparttable}
\begin{tabular}
{p{0.7cm} p{0.5cm} p{0.5cm} p{0.5cm} p{0.5cm} p{0.5cm} p{4.5cm} p{2cm} p{2cm} p{2cm} p{2cm} p{0.9cm}
}
 \multicolumn{8}{c}{} \\
 \hline \hline & &  \\ 
 Ion & & & & & & Pair of transitions & \multicolumn{4}{c}{Non-linearity (Hz)}\\ [0.3cm]
  & Z & A & A$_{1} $& A$_{2}$ & A$_{3}$ & & Method 4 & Method 5 &\multicolumn{2}{l}{Quadratic term inc. MB} & QMS\\
  \multicolumn{9}{c}{} & without $\alpha_p$ & with $\alpha_p$ \\
\hline
& &  \\ [-0.5cm]
Ca$^{+}$ & 20 & 40 & 42 & 44 & 48 & $3p^{6}4s\;\;^{2}S_{1/2}\rightarrow 3p^{6}3d\;\;^{2}D_{3/2}$ &$3.0\times 10^{-4}$ & $-6.6\times 10^{-2}$ & $\pm\ 
2.9 \times 10^{-3}$ & $\pm\ 2.7\times 10^{-3}$ & $\pm\ 3.0$ \\
& & & & & & $3p^{6}4s\;\;^{2}S_{1/2}\rightarrow 3p^{6}3d\;\;^{2}D_{5/2}$ & & & &\\
\hline
Sr$^{+}$ & 38 & 84 & 86 & 88 & 90 &  $4p^{6}5s\;\;^{2}S_{1/2}\rightarrow 4p^{6} 4d\;\; ^{2}D_{3/2}$ & $1.1 \times 10^{-2}$ & $-2.6$ & $\pm\ 
0.23$ & $\pm\ 0.25$ & $ \pm\ 9.0 $\\
& & & & & & $4p^{6}5s\;\;^{2}S_{1/2}\rightarrow 4p^{6} 4d\;\;^{2}D_{5/2}$ & & \\
\hline
Ba$^{+}$ & 56 & 132 & 134 & 136 & 138 & $ 5 p^6 6 s^1 \;\;^{2}S_{1/2}\rightarrow 5 p^6 5 d \;\;^{2}D_{3/2}$ & $-3.9 \times 10^{-2}$ & $7.4 $ & $\mp\ 2.0 $& $\mp\ 1.9 $ & $\mp\ 1.8$ \\
& & & & & & $ 5 p^6 6 s^1 \;\;^{2}S_{1/2}\rightarrow 5 p^6 5 d \;\;^{2}D_{5/2}$ & & \\
\hline
Yb$^{+}$ & 70 & 168 & 170 & 172 & 176 & $4f^{14}6s\;\;^{2}S_{1/2}\rightarrow 4f^{13}6s^{2}\;\; ^{2}F_{7/2}^{o}$ & $-3.1$ & $39$ & 
$\pm\ 12260$ & $\pm\ 12130$ & $\pm\  28 $\\ 
& & & & & & $4f^{14}6s\;\;^{2}S_{1/2}\rightarrow 4f^{14}5d\;\;^{2}D_{3/2}$ & & \\
\hline
& & & & & & $4f^{14}6s\;\;^{2}S_{1/2}\rightarrow 4f^{14}5d\;\;^{2}D_{3/2}$ & $3.1$ & $-18$ & 
$\pm\ 392$ & $\pm\ 386$ & $\pm\ 1.1 $ \\
& & & & & & $4f^{14}6s\;\;^{2}S_{1/2}\rightarrow 4f^{14}5d\;\;^{2}D_{5/2}$ & & \\
\hline
Hg$^{+}$ & 80 & 196 & 198 & 200 & 204 & $5d^{10}6s\;\; ^{2}S_{1/2}\rightarrow 5d^{9}6s^{2}\;\;^{2}D_{3/2}$ & $3.0$ & $-13$ & 
$\pm\ 2406$ & $\pm\ 2382$ & $\pm\ 0.38$\\
& & & & & & $5d^{10}6s\;\;^{2}S_{1/2}\rightarrow 5d^{9}6s^{2}\;\;^{2}D_{5/2}$\\
\hline \hline 
\end{tabular}

\end{threeparttable}
\end{table*}

\subsection{Many-body corrections}\label{sec:MB}
The results above have been obtained in the mean-field approximation. However, the isotope shifts in all waves with $|\kappa|>1$ are dominated by the many-body effects. In the zeroth approximation, the change of the isotope changes $s$~
and $p_{1/2}$ electron wave functions which do not vanish at the nucleus. This produces the correction $\delta V$ to the electron potential which gives the dominating contribution to the isotope shifts of the orbitals with $|\kappa|>1$; we will refer to this as the mean-field rearrangement effect. Therefore, in any wave the dominating term in the isotope shift  is proportional to  
$\delta \left < r^{2 \gamma_{1}} \right >_{AA'}$ corresponding to $|\kappa|=1$. However, the term with 
$\delta \left < r^{2 \gamma_{2}} \right >_{AA'}$ still appears in the transition frequencies and the logic of the section above does not change. The many-body corrections only affect the magnitude of the coefficients $F_i$ and $G_i$ in the isotope shift (see Eq.~\eqref{eq:mu}).
 
Consider, for example, King plot for  $s-d_{3/2}$ and $s-d_{5/2}$ transitions for Ca$^+$, Sr$^+$ and Yb$^+$  presented in the Table \ref{table:king}. Firstly, there are higher-order terms in the expansion of the $s$-wave density near origin, $\sim r^{2 \gamma_{1}+2}$. They only lead to the redefinition of the main term $\delta \left < r^{2 \gamma_{1}} \right >_{AA'}$ and do not produce any new physical effects. The mean-field rearrangement effect and other many-body corrections for $s$ orbital are relatively small, $\sim 10-20 \%$ and  give  contributions to the coefficient $F_i$ which are the same for both transitions and therefore insignificant.  The mean-field rearrangement effects for  $d_{3/2}$ and $d_{5/2}$ are huge in comparison with the direct contributions, but their absolute values are smaller than that for $s_{1/2}$. These mean-field rearrangement  effects produce some corrections to the coefficients $F_i$  and $G_i$ but do not give a significant contribution to the non-linearity of the King plot. This non-linearity comes from the direct contribution to the term  $G_i \delta \left < r^{2 \gamma_{2}} \right >_{AA'}$ since the density of $d_{3/2}$ orbital ($\kappa=2$) near nucleus is several orders of magnitude larger than the density of $d_{5/2}$  orbital ($\kappa=-3$). We have already taken this effect into account at the single-particle mean-field level. The IS correction to the potential  $\delta V$ is located at larger distances where the densities of  $d_{3/2}$ and $d_{5/2}$ are approximately the same.
 
 The nonlinear corrections may be significantly larger in atoms with several valence electrons. The density of energy levels is much higher in such systems. In this case the second order effects in the field shift  perturbation $\delta V$ in Eq. (\ref{eq:dv}) may be enhanced by small energy denominators and produce large non-linear effects in the King plot \cite{griffith_anomalies_1981,palmer_theory_1982,seltzer_$k$_1969,blundell_reformulation_1987,torbohm_state-dependent_1985}. 
 
We want to provide a rough numerical estimate of the many-body effects to the non-linearity of the King plot using a simple model. 
We have compared our mean field approximation for $s$ and higher waves with the accurate numerical CI+MBPT calculations used throughout \cite{berengut_probing_2017,berengut_isotope-shift_2003,berengut_private_nodate}. Many-body correction to the $s$-wave field shift is not very large \cite{berengut_isotope-shift_2003}. For the purpose of this work we can omit this term. \\
However, many body corrections for higher wave terms are found to be significant \cite{berengut_isotope-shift_2003}. For the higher waves $\delta V$ is dominated by the corrections to the $s$-wave functions, therefore we model the mean-field rearrangement effect by the following expression: 

\begin{equation}
\Delta\tilde{\varepsilon}_{\kappa}=\Delta\varepsilon_\kappa - \frac{\Delta\varepsilon_{s,\kappa}}{2}, \quad \Delta\varepsilon_{s,\kappa}=\Delta\varepsilon_s\left(\frac{I_\kappa}{I_s}\right)^{3/2}. \label{eq:kappa_mb1}
\end{equation}
Here the initial $\Delta\varepsilon_s=\Delta E_s+\delta E_{\alpha,s}$ comprises both mean-field contribution \eqref{eq:DE} and the polarisability term \eqref{eq:dEa_1}. The ratio of ionisation potentials comes from the density in the vicinity of the nucleus which is proportional to $I^{3/2}$ (see Eqs. \eqref{eq:g_bessel}-\eqref{defineC}). We have tested this semi-empirical estimate by comparing our field shift results with accurate numerical many-body calculations \cite{berengut_probing_2017,berengut_private_nodate}; the coefficient of $1/2$ was necessary to reproduce the numerical results. It must be noted that this approximation works best for alkali-like ions such as Ca$^+$ and Sr$^+$ and is less effective in characterising many-body effects in Yb$^+$ and Hg$^+$; in these heavier ions we expect an order of magnitude estimate of many-body effects at best.    

Non-linear corrections to the King plot may also be produced by the quadratic effects in the field shift, which we further estimate as 
$\pm \left(\Delta\varepsilon_\kappa - \frac{\Delta\varepsilon_{s,\kappa}}{2}\right)^2/I_\kappa$ \footnote{The quadratic effect may be enhanced if there is a close atomic level with the same angular momenta and parity which may be admixed by the IS operator. This does not happen in atoms with one electron above close shells which we consider.}. Therefore the complete formulae for the shift of an energy level will look as:
\begin{equation}
\Delta\tilde{\tilde{\varepsilon}}_{\kappa}=\Delta\tilde{\varepsilon}_\kappa \pm \frac{\left(\Delta\tilde{\varepsilon}_{\kappa}\right)^2}{I_\kappa}\ . \label{eq:kappa_mb2}
\end{equation}
Here $I_s$ denotes the ionization potential of an s-wave, $I_\kappa$ is that for any other wave. The quadratic effects arising from the s-wave can be similarly written as:
\begin{align}
\Delta\tilde{\tilde{\varepsilon}}_s&=\Delta{\varepsilon}_s \pm \frac{\left(\Delta{\varepsilon}_s\right)^2}{I_s}\ .\label{eq:s_mb2}
\end{align}
However in this $s$-wave formula we have omitted the many-body correction term which does not play an important role here. 


King plot non-linearity values taking into account many-body contributions are presented in Table \ref{table:kingMB}. As can be seen, the addition of the first order corrections 
\eqref{eq:kappa_mb1} does not considerably affect the non-linearity. 
It should be noted that for the evaluation of many-body effects one must include all effects which have been included at the mean-field level. For example, if the field shift term \eqref{eq:DE} and the nuclear polarizability contribution \eqref{eq:dEa_1} have been included at the mean-field level they both must be included in the mean-field rearrangement and quadratic effects. If it contains only the first term, the resulting first and second order many-body effects
 will generate a large phantom non-linearity in the King plot, because the dependence on the nuclear parameters (radius $R$ and mass number $A$) is no longer the same for the single-electron mean-field and many-body effect.

On the other hand, the quadratic term 
\eqref{eq:kappa_mb2} is responsible for the radical growth of King plot non-linearity in heavy atoms (especially for f-shell transition in Yb$^{+}$), while remaining insignificant in Ca$^+$ and Sr$^+$.
Indeed, the field shift is $\propto Z^{2\gamma}$, correspondingly the quadratic term is $\propto Z^{4\gamma}$, i.e. it very rapidly increases with the nuclear charge.

\section{Estimate of the quadratic mass shift in the King plot non-linearity}  
For one electron above closed shells, the normal mass shift may be used as a rough estimate for the total mass shift \cite{berengut_isotope-shift_2003,safronova_third-order_2001}: 

\begin{align}
\Delta{\varepsilon}_{M}&= -\varepsilon m_e  \mu_{AA'}\ , \label{eq:M} \\
\Delta\tilde{\tilde{\varepsilon}}_{M2}&=\pm \frac{\left(\Delta{\varepsilon}_M\right)^2}{I}\ ,\label{eq:M2}
\end{align}
where $m_e$ is the electron mass, $\varepsilon$ is the energy of a specific electronic level and $\Delta \varepsilon$ is the shift in energy of this level. The non-linearity of the King plot (see Eq. \eqref{eq:nonlin_def}) arising from the addition of the quadratic term \eqref{eq:M2} to the IS is shown in the last column of Table \ref{table:kingMB}. The linear normal mass shift term \eqref{eq:M} does not contribute to the non-linearity.

\section{New particle}
King plot non-linearity 
may result from an interaction between electrons and neutrons mediated by a new boson of mass $m_\phi$ 
\cite{delaunay_probing_2017}. The effective potential associated with such a particle would be the Yukawa potential:
\begin{equation}\label{eq:yukawa}
V_\phi(r)=-q_n q_e N\frac{e^{-kr}}{r}\ ,
\end{equation}
\begin{equation*}
k=\frac{m_\phi c}{\hbar},\quad \alpha_\mathrm{NP} \equiv\frac{q_n q_e}{\hbar c}\ ,
\end{equation*}
here $N$ is the neutron number, $q_n$ and $q_e$ are particle coupling strengths to the neutrons and electrons respectively, $r$ is the distance from the nucleus. We aim at constraining the coupling constant $\alpha_\mathrm{NP}$. Let us estimate the energy shifts in atomic states that the new particle might cause. When the particle is very light
 ($k \ll 1/a_B$) and therefore $e^{-kr}\approx 1$, the potential \eqref{eq:yukawa} becomes Coulomb-like:
\begin{equation}
V_\phi(r)=- q_n q_e N\frac{1}{r}\ .
\end{equation}
Making use of the virial theorem, one can express the average potential energy of the system as double total energy:
\begin{equation}
\left<V\right>=2 E_\mathrm{tot}\ .
\end{equation}
Substituting here values for a single outer electron in the Coulomb field 
 $V=V_c=-(z_i+1)e^2/r$ and $E_\mathrm{tot}=-I_\kappa$, one obtains:
\begin{equation}
\left<\frac{1}{r}\right>= \frac{2I_\kappa}{(z_i+1) e^2}\ .
\end{equation}
Therefore the energy shift of an electron with a given $\kappa$ arising from a new light particle, seen as the change of $\left<V_\phi \right>$ between the isotopes, can be approximately written as:
\begin{equation}\label{eq:lightphi}
\Delta E_{\phi,\kappa}=-\frac{\alpha_\mathrm{NP}}{\alpha}\frac{ 2 I_\kappa}{(z_i+1)} \Delta N\ .
\end{equation}
We replaced $q_n q_e/e^2=\alpha_\mathrm{NP}/\alpha$, where $\alpha=e^2/\hbar c$ is fine structure constant. For mass $m_{\phi}=0$ many-body effects are not enhanced, so we do need to add them.

On the other hand, the energy shifts resulting from an interaction mediated by heavier particles 
($Z^{1/3}/a_B< k<1/R$, here $R$ is the nuclear radius, and $a_B/Z^{1/3}$ is the Thomas-Fermi electron screening radius) are found by direct integration of \eqref{eq:yukawa} with the relativistic wave functions for a valence electron (see Appendix B):
\begin{equation}\label{eq:heavyphi}
\Delta E_{\phi,\kappa} = - q_n q_e \Delta N\int_0^\infty \left(f^2_\kappa(r)+g^2_\kappa(r)\right)\frac{e^{-kr}}{r}r^2dr\ . 
\end{equation}
For larger masses, $Z/a_B< k<1/R$,    it is instructive to present also approximate formula for \eqref{eq:heavyphi} which shows dependence on the Compton wavelength $k=\frac{m_\phi c}{\hbar}$ and other parameters explicitly:
\begin{align}
\Delta E_{\phi,\kappa} = -\frac{\alpha_\mathrm{NP}}{\alpha}\frac{4\kappa(\kappa-\gamma)}{(z_i+1)Z}\frac{\Gamma(2\gamma)}{[\Gamma(2\gamma+1)]^2}\left(\frac{2 Z}{k a_B}\right)^{2\gamma}
\nonumber\\ \times
\frac{I_\kappa^{3/2}}{\mathrm{Ry}^{1/2}}\Delta N\ .
\end{align}

The effect of very heavy bosons with $k>1/R$ is absorbed into the usual field shift (proportional to $R^{2\gamma}$), as the range of the interaction is less than the nuclear radius. It is therefore nearly impossible to see this new physics effect against the background of nuclear uncertainties. We restrict our consideration with masses corresponding to $k<1/R$, i.e. $m_\phi\lesssim 30$~MeV.

We also take into account the characteristic value of many-body mean-field rearrangement corrections. We estimate these many-body effects in a similar manner to the corrections to the isotope shift seen in (\ref{eq:kappa_mb1}). The mean-field rearrangement effect can be modeled as:

\begin{equation}
\Delta\tilde{E}_{\phi,\kappa}=\Delta E_{\phi,\kappa} - \frac{\Delta E_{\phi,s,\kappa}}{2}, \quad \Delta E_{\phi,s,\kappa}=\Delta E_{\phi,s} \left(\frac{I_\kappa}{I_s}\right)^{3/2}. \label{eq:Yukawa_kappa_mb1}
\end{equation}

The coefficient $1/2$ arises from comparisons of the electronic part of the integral (\ref{eq:heavyphi}) (omitting $\Delta N$) with accurate numerical many-body calculations of the same quantity performed in \cite{berengut_probing_2017}. Many-body effects are taken into account for higher waves only.

As always, the IS of a transition frequency would be the difference of the shifts of two levels $\Delta E_{\phi,\kappa}$. We examine the non-linearity arising from the addition of these terms to otherwise linear King plot.  Firstly, we construct the linear King plot leaving  only the  mean field  $s$-wave contribution of the form \eqref{eq:DE} in all transitions and then we add the new particle contribution to this linear King plot.Then we examine the sensitivity of non-linearity to the coupling constant $\frac{\alpha_\mathrm{NP}}{\alpha}$ by equating the non-linearity which appears as a result of including a new particle and the non-linearity emerging naturally from SM. The values of $\frac{\alpha_\mathrm{NP}}{\alpha}$ that lead to the same non-linearity as SM 
 corrections in a given pair of transitions are presented in Table \ref{tab:alpha_NP}. 

\begin{table*}[htb]
\caption{\label{tab:alpha_NP} Values of the ratio $\frac{\alpha_\mathrm{NP}}{\alpha}$ of the coupling constant $\alpha_\mathrm{NP}$ for a new boson with mass $m_\phi$ to the fine structure constant $\alpha$.
 They correspond to the most significant King plot non-linearity value that can arise from SM 
 corrections to the IS (see Table \ref{table:kingMB}). Individual isotope shifts consist of the s-wave contribution, which does not give rise to any non-linearity, and the new particle contributions \eqref{eq:lightphi} and \eqref{eq:heavyphi} for light ($m_\phi\rightarrow 0$) and heavier particles respectively.}~\\[-0.5cm]
 \begin{threeparttable}
\begin{tabular}
{p{0.7cm} p{0.5cm} p{0.5cm} p{0.5cm} p{0.5cm} p{0.5cm} p{4.5cm} p{2.0cm} p{1.85cm} p{1.85cm} p{1.85cm} p{1.8cm}
}
 \multicolumn{8}{c}{} \\
 \hline \hline & &  \\ 
 Ion & & & & & & Pair of transitions & Non-linearity (Hz) & $\frac{\alpha_\mathrm{NP}}{\alpha}$  \\ [0.3cm]
  & Z & A & A$_{1} $& A$_{2}$ & A$_{3}$ & & & \footnotesize{$m_\phi\rightarrow 0$} & \footnotesize{$m_\phi=10^5$~eV} & \footnotesize{$m_\phi=10^6$~eV} & \footnotesize{$m_\phi=10^7$~eV} \\
\hline
& &  \\ [-0.5cm]
 Ca$^{+}$ & 20 & 40 & 42 & 44 & 48 & $3p^{6}4s\;\;^{2}S_{1/2}\rightarrow 3p^{6}3d\;\;^{2}D_{3/2}$ & $-3.0$ & $5.6\times 10^{-12}$ & $1.4\times 10^{-9}$ & $-9.9\times 10^{-9}$ & $-7.2\times 10^{-7}$\\
 & & & & & & $3p^{6}4s\;\;^{2}S_{1/2}\rightarrow 3p^{6}3d\;\;^{2}D_{5/2}$ & & & \\
 \hline
 Sr$^{+}$  & 38 & 84 & 86 & 88 & 90 &  $4p^{6}5s\;\;^{2}S_{1/2}\rightarrow 4p^{6} 4d\;\; ^{2}D_{3/2}$ & $-11.9$ & $6.7\times 10^{-13}$ & $5.5\times 10^{-11}$ & $-7.4\times 10^{-10}$ & $-3.8\times 10^{-8}$\\
  & & & & & & $4p^{6}5s\;\;^{2}S_{1/2}\rightarrow 4p^{6} 4d\;\;^{2}D_{5/2}$ & \\
  \hline
Ba$^{+}$ & 56 & 132 & 134 & 136 & 138 & $ 5 p^6 6 s^1 \;\;^{2}S_{1/2}\rightarrow 5 p^6 5 d \;\;^{2}D_{3/2}$ & 11.1 & $7.7 \times 10^{-13}$ & $3.9 \times 10^{-11}$ & $-3.4 \times 10^{-8}$ & $-3.3 \times 10^{-7}$ \\
& & & & & & $ 5 p^6 6 s^1 \;\;^{2}S_{1/2}\rightarrow 5 p^6 5 d \;\;^{2}D_{5/2}$ & & \\
\hline

 Yb$^{+}$ & 70 & 168 & 170 & 172 & 176 & $4f^{14}6s\;\;^{2}S_{1/2}\rightarrow 4f^{13}6s^{2}\;\; ^{2}F_{7/2}^{o}$ & $12190$ & $-2.5\times 10^{-11}$ & $2.4\times 10^{-9}$ & $1.3\times 10^{-8}$ & $3.2\times 10^{-7}$ \\ 
  & & & & & & $4f^{14}6s\;\;^{2}S_{1/2}\rightarrow 4f^{14}5d\;\;^{2}D_{3/2}$ & \\
  \hline
  & & & & & & $4f^{14}6s\;\;^{2}S_{1/2}\rightarrow 4f^{14}5d\;\;^{2}D_{3/2}$ & $-406$ & $2.7\times 10^{-11}$ & $2.2\times 10^{-9}$ & $-1.7\times 10^{-8}$ & $-3.9\times 10^{-7}$ \\
  & & & & & & $4f^{14}6s\;\;^{2}S_{1/2}\rightarrow 4f^{14}5d\;\;^{2}D_{5/2}$ & \\
  \hline
  Hg$^{+}$ & 80 & 196 & 198 & 200 & 204 & $5d^{10}6s\;\; ^{2}S_{1/2}\rightarrow 5d^{9}6s^{2}\;\;^{2}D_{3/2}$ & $-2395$ & $-1.8\times 10^{-10}$ & $6.6\times 10^{-8}$ & $-5.5\times 10^{-8}$ & $-1.0\times 10^{-6}$ \\
  & & & & & & $5d^{10}6s\;\;^{2}S_{1/2}\rightarrow 5d^{9}6s^{2}\;\;^{2}D_{5/2}$\\
  \hline \hline 
\end{tabular}

\end{threeparttable}
\end{table*}

\section{Conclusions}
The analytical formula for the field isotope shift \eqref{eq:dE} gives reasonable accuracy of the estimates for the transitions involving $s$-wave electron. In superheavy elements it should also describe the $p_{1/2}$-wave field IS which is comparable to the $s$-wave shift. For higher waves the field IS is dominated by the many-body corrections which 
are in turn dominated by the mean-field rearrangement effect. The latter is produced by the IS of the mean field potential due to IS of the $s$-electron wave functions. 

In the single-particle mean-field approximation the non-linearity of the King plot is strongly dominated by the nuclear polarizability contribution 
(see Table \ref{table:king}). However, the quadratic field shift, which very rapidly increases with the nuclear charge $Z$ and gives the dominating contribution to the non-linearity of the King plot in heavy atoms, is not so sensitive to the nuclear polarizability contribution (see Table \ref{table:kingMB}). However, in medium  atoms the quadratic terms are not so large, therefore, the measurements of the non-linearity of the King plot may, in principle, be used to extract the nuclear polarizability differences between the isotopes.
  
The contribution of the hypothetical new light boson increases with the nuclear charge $Z$. However, the quadratic contribution to  the field IS increases with $Z$ much faster. In light atoms the non-linearity is dominated by the quadratic mass shift. Therefore, it may be easier  to extract a competitive limit on the new particle interaction strength from the measured non-linearity of the King plot in medium atom transtions such as $s-d$ transitions in Sr$^+$ and Yb$^+$  -  see Table \ref{tab:alpha_NP}.
  
  The field IS may be  an order of magnitude smaller in transitions which do not involve $s$-wave electrons. This means that the dominating source of the King plot non-linearity in heavy atoms, the quadratic field IS term, may be much smaller. Such transitions may, in principle, provide better accuracy for the low mass new particle. However, these must be transitions with a small natural width. In all existing optical atomic clocks such narrow transitions always involve $s-$electron.  Therefore, to explore such possibility we should look for  narrow transitions in atoms and ions containing $p$, $d$ or $f$ electrons in the ground open shell, or low energy excitations from the closed $f$, $d$, $p$ shells.

\section{Acknowledgements}
This work was supported by the Australian Research Council and the Gutenberg Fellowship.
The authors are grateful to S. Karshenboim,  D. Budker, K. Pachucki, M. Pospelov, V. Shabaev and E. Fuchs for valuable discussions and corrections. A.V.V. would like to thank UNSW, Australia for hospitality.

\section*{Appendix}
\subsection{Relativistic electron wave function inside nucleus}
The $s$ and $p_{1/2}$ approximate  wave functions within the nucleus in a neutral atom are presented in \cite{flambaum_nuclear_2002,khriplovich_parity_1991}. We have introduced minor changes to extend the result to ions with charge $z_i$. Denoting $x=r/R$, the upper and lower radial components of the $s$ wave function can be given:
\begin{equation}
f_{s} = A_{s} \left\lbrack 1-\frac{3}{8} Z^{2}\alpha^{2} x^2 \left( 1- \frac{4}{15} x^2\right) \right\rbrack\ ,
\end{equation}
\begin{align}
g_{s} = & -\frac{1}{2} A_{s} Z \alpha x \nonumber \\ &\times \left \lbrack 1 - \frac{1}{5} x^2 -\frac{9}{40} Z^{2}\alpha^{2}  x^2 \left( 1 - \frac{3}{7} x^2 + \frac{4}{81}x^4 \right)  \right \rbrack\ ,
\end{align}
Where $A_{s}$ is a constant defined as
\begin{align}
A_{s} = \frac{2}{(z_i+1)^{1/2}}\frac{2 (\frac{a_B}{2ZR})^{1-\gamma}}{\Gamma(2 \gamma + 1)} \left( \frac{Z}{a_B^{3}} \right )^{1/2}\nonumber \\ \times \left( \frac{I}{\mathrm{Ry}} \right )^{3/4} \left( 1 - \frac{1}{40} Z^{2}\alpha^{2} \right).
\end{align}
Here  $\gamma = \sqrt{\kappa^{2} - Z^{2}\alpha^{2}}$,  $I=\frac{(z_i+1)^2}{\nu^2}\mathrm{Ry}$ is the ionization energy with effective principal quantum number $\nu$ and $\mathrm{Ry}=\frac{e^{2}}{2 a_B}$. 

The radial $p_{1/2}$ wave functions are written in terms of the radial $s$ wave functions in the following way:

\begin{align}
f_{p}&=-\frac{A_{p}}{A_{s}}\;g_{s}\ ,\\
g_{p}&=i \frac{A_{p}}{A_{s}}\;f_{s}\ .
\end{align}

Here $ A_{p} $ is
\begin{align}
A_{p} = \frac{Z\alpha}{(z_i+1)^{1/2}} \frac{2 (\frac{a}{2ZR})^{1-\gamma}}{\Gamma(2 \gamma + 1)} \left( \frac{Z}{a_B^{3}} \right )^{1/2} \nonumber\\ \times \left( \frac{I}{\mathrm{Ry}} \right )^{3/4}  \left( 1 + \frac{9}{40} Z^{2}\alpha^{2} \right).
\end{align}

\subsection{Relativistic wave function for a valence electron at  $r\ll a_B/Z^{1/3}$}
 At short distances $r\ll a_B/Z^{1/3}$ the nuclear Coulomb potential is not screened and the valence electron energy may be neglected. The solution of the Dirac equation can be expressed in terms of Bessel functions - see e.g. \cite{khriplovich_parity_1991}.
\begin{align}
f_\kappa &=\frac{C}{r}\left[(\gamma+\kappa)J_{2\gamma}(y)-\frac{y}{2}J_{2\gamma-1}(y)\right] \label{eq:f_bessel}\\
g_\kappa &=\frac{C}{r}\left(Z\alpha\right)J_{2\gamma}(y) \label{eq:g_bessel}
\end{align}
\begin{equation}
y=\sqrt{\frac{8Zr}{a_B}}
\end{equation}
\begin{equation}
C=\frac{\kappa}{|\kappa|}\frac{1}{\sqrt{Za_B(z_i+1)}}\left(\frac{I}{\mathrm{Ry}}\right)^{3/4}
\label{defineC}
\end{equation}
Again, we introduced factor $(z_i+1)$ to account for ion wave functions.

\subsection{Relativistic wave function for a valence electron at  $r\ll a_B/Z$}
The Bessel functions have power asymptotic at  $r\ll a_B/Z$. From power expansions of \eqref{eq:f_bessel} and \eqref{eq:g_bessel}
one obtains the following expression for the electron wave functions outside the nucleus \cite{khriplovich_parity_1991,sobelman_introduction_1972}:
\begin{align}
f_\kappa(r)&=\frac{1}{(z_i+1)^{1/2}}\frac{\kappa}{|\kappa|}(\kappa-\gamma)\left(\frac{Z}{a_B^3}\right)^{1/2} \\ \nonumber & \times \left( \frac{I}{\mathrm{Ry}} \right )^{3/4}\frac{2}{\Gamma(2\gamma+1)}\left(\frac{a_B}{2Zr}\right)^{1-\gamma} \\
g_\kappa(r)&=\frac{1}{(z_i+1)^{1/2}}\frac{\kappa}{|\kappa|}Z\alpha \left(\frac{Z}{a_B^3}\right)^{1/2} \\ \nonumber & \times \left( \frac{I}{\mathrm{Ry}} \right )^{3/4}\frac{2}{\Gamma (2\gamma+1)}\left(\frac{a_B}{2Zr}\right)^{1-\gamma}
\end{align}

\subsection{Ionization potentials for isotope shift and King plot calculations}
The ionization potentials of the $7s$ electrons for the superheavy elements $Z=98-102$ are taken from the NIST database \cite{kramida_nist_2017}. Those include experimental values 50665~cm$^{-1}$ and 51358~cm$^{-1}$ for $Z=98,99$, semiempirical evaluations 52400~cm$^{-1}$ and 53100~cm$^{-1}$ ($Z=100,101$) and a theoretical calculation 53740~cm$^{-1}$ ($Z=102$). The ionization potential 59462~cm$^{-1}$ of Rf ($Z=104$) is roughly estimated as the average of potentials for $Z=102$ and $Z=105$ due to the lack of reliable information.

The energies for $7p$ electrons in $Z=98,99,102$ are derived from the experimental values of the relevant transitions 27779~cm$^{-1}$, 19788~cm$^{-1}$ and 29961~cm$^{-1}$ \cite{sansonetti_handbook_2005,laatiaoui_atom-at--time_2016} and in ($Z=104$) the energy of the transition 20347~cm$^{-1}$ is based on a numerical prediction \cite{dzuba_atomic_2014}. For Fm ($Z=100$) and Md ($Z=101$) the $7p$ ionization potential is taken to be the average of known Es ($Z=99$) and No ($Z=102$) potentials, 27675~cm$^{-1}$.

For elements $Z=105-112$ we use numerical values of $7s$ and $7p$ ionization potentials \cite{dzuba_private_nodate}.


Furthermore, the $p$ ground state ionization energy in Lr is measured to be 
40005~cm$^{-1}$ \cite{sato_measurement_2015}. The upper $s$ state was found by subtracting the $p\rightarrow s$ calculated transition frequency of 20253~cm$^{-1}$ \cite{dzuba_atomic_2014} 
to give an ionization potential of 
19800~cm$^{-1}$. Recent atomic structure calculations were used to find the potentials for the ground $p$ states and excited $s$ states for Nh and Fl. For the Nh $p$ state we used 
59770~cm$^{-1}$
and the $s$ state ionisation potential of 
23729~cm$^{-1}$ \cite{dinh_all-order_2016}. Similarly, for the Fl $p$ state we used 68868~cm$^{-1}$ \cite{landau_electronic_2001}, a $p \rightarrow s$ transition energy of 
43876~cm$^{-1}$ \cite{dinh_all-order_2016} to give a $s$ state ionisation potential of 24992~cm$^{-1}$.

Ionization potentials and transition energies for King plot non-linearity estimates were taken from the NIST database \cite{kramida_nist_2017}.
\bibliographystyle{apsrev}

\end{document}